\documentclass[twocolumn,showpacs,preprintnumbers,amsmath,amssymb,superscriptaddress,amssymb]{revtex4-2}
\usepackage{subcaption}
\usepackage{graphicx}
\usepackage{epsfig}
\usepackage{multirow}
\usepackage{pbox}
\usepackage{array}
\usepackage{relsize}
\usepackage{siunitx}
\usepackage{braket}
\usepackage[labelfont=normalfont,textfont=normalfont,singlelinecheck=off,justification=justified]{caption}
\UseRawInputEncoding
\begin{document}
\newcolumntype{P}[1]{>{\centering\arraybackslash}p{#1}}
\title{A global study of  $^9$Be + p at 2.72 A MeV}
\author{V. Soukeras}
\affiliation{INFN Laboratori Nazionali del Sud, via S. Sofia 62, 95125, Catania, Italy}
\author{O. Sgouros}
\affiliation{INFN Laboratori Nazionali del Sud, via S. Sofia 62, 95125, Catania, Italy}
\author{A. Pakou}
\email{apakou@uoi.gr}
\affiliation{Department of Physics and HINP, The University of Ioannina, 45110 Ioannina,
Greece}
\author{F. Cappuzzello}
\affiliation{INFN Laboratori Nazionali del Sud, via S. Sofia 62, 95125, Catania, Italy}
\affiliation{Dipartimento di Fisica e Astronomia "Ettore Majorana", Universit\`{a} di Catania, via S. Sofia 64, 95125, Catania, Italy}
\author{J. Casal}
\affiliation{Dipartimento di Fisica e Astronomia "G. Galilei", Universit\`{a}  degli Studi di Padova, Via Marzolo 8, I-35131 Padova, Italy}
\affiliation{INFN - Sezione di Padova, Via Marzolo 8, I-35131 Padova, Italy}
\author{C. Agodi}
\affiliation{INFN Laboratori Nazionali del Sud, via S. Sofia 62, 95125, Catania, Italy}
\author{G. A. Brischetto}
\affiliation{INFN Laboratori Nazionali del Sud, via S. Sofia 62, 95125, Catania, Italy}
\affiliation{Dipartimento di Fisica e Astronomia "Ettore Majorana", Universit\`{a} di Catania, via S. Sofia 64, 95125, Catania, Italy}
\author{S. Calabrese}
\affiliation{INFN Laboratori Nazionali del Sud, via S. Sofia 62, 95125, Catania, Italy}
\affiliation{Dipartimento di Fisica e Astronomia "Ettore Majorana", Universit\`{a} di Catania, via S. Sofia 64, 95125, Catania, Italy}
\author{D. Carbone}
\affiliation{INFN Laboratori Nazionali del Sud, via S. Sofia 62, 95125, Catania, Italy}
\author{M. Cavallaro}
\affiliation{INFN Laboratori Nazionali del Sud, via S. Sofia 62, 95125, Catania, Italy}
\author{I. Ciraldo}
\affiliation{INFN Laboratori Nazionali del Sud, via S. Sofia 62, 95125, Catania, Italy}
\affiliation{Dipartimento di Fisica e Astronomia "Ettore Majorana", Universit\`{a} di Catania, via S. Sofia 64, 95125, Catania, Italy}
\author{I. Dimitropoulos}
\affiliation{Department of Chemistry, National and Kapodistrian University of Athens and HINP, 15771 Athens,
Greece}
\author{S. Koulouris}
\affiliation{Department of Chemistry, National and Kapodistrian University of Athens and HINP, 15771 Athens,
Greece}
\author{L. La Fauci}
\affiliation{INFN Laboratori Nazionali del Sud, via S. Sofia 62, 95125, Catania, Italy}
\affiliation{Dipartimento di Fisica e Astronomia "Ettore Majorana", Universit\`{a} di Catania, via S. Sofia 64, 95125, Catania, Italy}
\author{I. Martel}
\affiliation{Departamento de Ciencias Integradas, Facultad de Ciencias Experimentales,
Campus de El Carmen, Universidad de Huelva, 21071, Huelva, Spain}
\author{M. Rodr\'{\i}guez-Gallardo}
\affiliation{Departamento de F\'{\i}sica At\'omica, Molecular y Nuclear, Facultad de F\'{\i}sica, Universidad de Sevilla, Apartado 1065, E-41080 Sevilla, Spain  }
\affiliation{Instituto Carlos I de F\'{\i}sica Te\'orica y Computacional, Universidad de Sevilla, Spain}
\author{A. M. S\'anchez-Ben\'{\i}tez}
\affiliation{Centro de Estudios Avanzados en F\'{\i}sica, Matem\'aticas y Computaci\'on (CEAFMC), Department of Integrated Sciences, University
of Huelva, 21071 Huelva, Spain}
\author{G. Souliotis}
\affiliation{Department of Chemistry, National and Kapodistrian University of Athens and HINP, 15771 Athens,
Greece}
\author{A. Spatafora}
\affiliation{INFN Laboratori Nazionali del Sud, via S. Sofia 62, 95125, Catania, Italy}
\affiliation{Dipartimento di Fisica e Astronomia "Ettore Majorana", Universit\`{a} di Catania, via S. Sofia 64, 95125, Catania, Italy}
\author{D. Torresi}
\affiliation{INFN Laboratori Nazionali del Sud, via S. Sofia 62, 95125, Catania, Italy}
\date{\today}
\begin{abstract}
{\bf Background :}  In our recent experiment, $^9$Be + p at 5.67 A MeV,  the breakup decay rates to the three configurations,  $\alpha$ + $\alpha$ + n, $^8$Be$^*$ + n and $^5$He + $^4$He of $^9$Be, were observed and quantified in the proton recoil spectra, in a full kinematics approach. Unfolding step by step the accessibility to the above configurations, it will require similar experiments at lower or/and higher energies. It will also require the interpretation of the data in a theoretical framework. Three-body models for the structure of $^9$Be have been developed and applied to reactions with heavy targets. Further research on lighter targets is required for the best establishement of the model. Such models are relevant for the calculation of the corresponding radiative capture reaction rate, $\alpha$($\alpha$, $\gamma$)$^9$Be followed by $^9$Be($\alpha$,n)$^{12}$C. The last is essential for the r-process abundance predictions. \par
{\bf Purpose :} Investigate the breakup decay rate of $^9$Be + p at 2.72 A MeV, where the direct configuration $\alpha$ + $\alpha$ + n is mainly accessible. Compare and interpret data at this low energy and  at the higher energy of 5.67 A MeV into a four - body Continuum Discretized Coupled Channel formalism; Point out and discuss couplings to continuum.\par
{\bf Methods :} Our experimental method includes an exclusive breakup measurement in a full kinematic approach of $^9$Be incident on a proton target at 24.5 MeV (2.72 A MeV). Complementary the elastic scattering is measured and other reaction channels are evaluated from previous measurements under the same experimental conditions.  The interpretation of present data at 2.72 A MeV and previous data at 5.67 A MeV, are considered in a four - body Continuum Discretized Coupled Channel approach, using the Transformed Harmonic Oscillator method for the three - body projectile.\par
{\bf Results :}  An elastic scattering angular distribution at 2.72 A MeV is measured, which compares very well with CDCC calculations, indicating a strong coupling to continuum. At the same energy, the breakup and total reaction cross sections are measured as  $\sigma$$_{break}$ = 2.5 $\pm$ 1 mb and  $\sigma$$_{tot}$ = 510 $\pm$ 90 mb, in good agreement with the calculated values of 3.7 mb and 433 mb, respectively. Further on, into the same theoretical framework the elastic scattering and breakup cross section data at  5.67 A MeV are found  in very good agreement with the CDCC calculations. The present results support further our three - body model for the structure of $^9$Be, validatting relevant radiative reaction rates obtained previously.\par
\end{abstract}
\pacs{25.70Mn, 24.10.Eq, 27.20.+n}
\maketitle
\vspace{0.2cm}
\vspace{0.3cm}

\section{INTRODUCTION }

The effect of continuum on the reaction dynamics of weakly bound nuclei is an open subject of ongoing interest, for probing nuclear structure, coupling mechanisms and for contributing on astrophysical problems  \cite{aa1,aa2,aa3,aa4,aa5,ro1,astro2d,astro2}. In this direction, lithium and beryllium isotopes play a substantial role. Therefore we have undertaken a broad experimental program with the involvement of these nuclei, on breakup and other reaction channels induced by protons, and their interpretation  at appropriate theoretical frameworks, in respect with couplings to continuum. Such studies at low energies are also usefull for providing in the community and the new generations to come, experimental fundamental data and accociated theoretical descriptions, for different applications from astrophysics and material science to 
 societal applications on health,
energy and climate \cite{general1,general2,general3,general4}. Global experimental studies, interpeted into appropriate theoretical models can provide the basis for such applications. 

Our first results, related with $^{6,7}$Li exhibiting a two body cluster structure ($\alpha$ +d (t)), include not only elastic scattering  \cite{pakou3,pakou4} and breakup measurements \cite{pakou5,pakou6}, but also cross sections of all involved reaction channels \cite{pakou7,pakou8}. In this respect a global description into a three - body Continuum Discretized Coupled Channel (CDCC ) framework was successfully obtained. Measurements were made at low energies $\approx$ 6 times the Coulomb barrier. Recently at similar energies, we have measured the elastic scattering and breakup of a three - body cluster structure nucleus ($\alpha$ + $\alpha$ + n), the Borromean nucleus $^9$Be, at 5.67 A MeV \cite{pakou9,pakou10,pakou11,pakou12}. Decay rates were determined for the 2.429 ( 5/2$^-$ ) MeV resonance of $^9$Be into three decay modes, $^5$He + $^4$He, $^8$Be + n and $\alpha$ + $\alpha$ + n, and a total breakup cross section was deduced. It was also found that the decay via the $^5$He + $^4$He configuration is the strongest one, in good agreement with 
 beta decay and transfer measurements \cite{ge,borge,cha,lu,a5,a6}, but not with inelastic scattering studies on light targets \cite{a1,a2,a3}. No theoretical interpretation of these data was performed. In an attempt to unfold step by step a complicated situation, where according to the available energy of the system the breakup can occur either directly or/and sequentially via $^9$Be resonances and in turn via one or more of the three clustering modes and their resonances, within this work, we will describe a measurement at the low energy of 2.72 A MeV, close to three times the Coulomb barrier. The available energy above threshold in this case is only 0.88 MeV (  E$_{c.m.}$ - E$_{thresh}$ = (2.45 - 1.57 ) MeV). This means that the breakup will occur either directly or via  the broad first resonance of $^9$Be at 1.684 MeV (1/2$^+$) with $\Gamma$ = 214 keV. The main involved configuration according to the available energy will be the $\alpha$ + $\alpha$ + n one. Another candidate could have been the ground state of $^8$Be + n with threshold energy 1.6 MeV but our setup prevents this observation. 

From the theoretical point of view, scattering of weakly bound nuclei at low energies, request a special treatement, beyond the standard microscopic and folded potential frameworks \cite{ala,sat}. Satchler and Love  were the first to suggest that for the scattering of weakly bound nuclei as $^6$Li, double folded potentials using the traditional M3Y interaction, adapted successfully for the description of stable projectiles, need to be renormalized by a factor of $\approx$ 2 \cite{sat2}. The origin of this strong reduction was later understood within  three-body coupled-channel CDCC theories by Sakuragi and co-workers \cite{sak1,sak2,sak3,sak4}. The history of development of three - body CDCC theories can be found between other articles in Ref. \cite{nun}.  Reaction studies with $^9$Be present a real challenge as they are not anymore a three - body but a four -  body problem. For that the  CDCC formalism was extended to a four - body formalism  in a Transformed Harmonic Oscillator framework (THO) \cite{MRG08,andd,ro1}. The new formalism has been successfully applied to several reactions induced by $^9$Be \cite{ro2,ro3}. Within the present study, we have to face an additional challenge,  moving  at a very low energy regime, where breakup is expected to be small and coupling mechanisms to other reaction channels may prove to be important \cite{pakou10}. Therefore to support our understanding via our CDCC calculations, we will explore experimentally other reaction channels which may be of major importance and could remove flux from the elastic scattering channel. These are :
 \begin{itemize}
 \item 
 The charge exchange  reaction :
 p + $^9$Be $\rightarrow$ $^9$B + n $\rightarrow$ $\alpha$ + $\alpha$ + p + n 
 
 \item
 The neutron stripping reaction :
 p + $^9$Be $\rightarrow$ $^8$Be + d $\rightarrow$ $\alpha$ + $\alpha$ + d 
 \item
 A reaction with intermediate nucleus, $^{10}$B$^*$, decaying to ground state as well as to the  first and second excited states of $^6$Li :
  p + $^9$Be $\rightarrow$ $^{10}$B$^*$ $\rightarrow$  $^6$Li$^*$ + $\alpha$

 \end{itemize}
  For these reactions and at similar energies to the present one, extensive angular distribution data exist in the literature \cite{rea1,rea2,rea3,rea4}, which can be integrated and reaction cross sections can be deduced. These results can be validated through our measurements and a total reaction cross section can be determined, and used to validate the calculation. It should be noted however, that our experimental setup was tuned for the breakup reaction, therefore the other reaction data will be limited to a narrow angular range. 
  
 Our theoretical framework will be also applied to higher energy elastic scattering data at 5.67 A MeV, collected under similar experimental conditions \cite{pakou9,pakou10,pakou11}. In this respect  our theory can be validated in a more global context and fruitful comparisons can be deduced. Further on, this will give more support to our three - body structure model for $^9$Be, used previously for radiative capture studies \cite{ro1}, relevant to r-process abundance predictions. In neutron rich environments, instead of the standard triple - $\alpha$ formation of $^{12}$C,  the radiative capture reaction $\alpha$($\alpha$, $\gamma$)$^9$Be followed by the $^9$Be($\alpha$, n)$^{12}$C one may dominate, depending on the astrophysical conditions \cite{astro2}. The relevance of this process has been linked to the nucleosynthesis by rapid neutron capture (or r process) in
type II supernovae \cite{astro2,astro2a,astro2b}. Therefore establishing an accurate rate for
the formation of $^9$Be is essential for the r-process abundance predictions \cite{astro2c,astro2d}.

Another issue of special importance to be considered in this study is the reciprocity between the strength of breakup probability and the one of the coupling mechanism to continuum.  In Ref. \cite{keeley1}, the authors investigate breakup coupling effects on near-barrier $^6$Li, $^7$Be and $^8$B + $^{58}$Ni elastic scattering in a CDCC approach. They observe the following paradox: $^6$Li, with a relatively small breakup cross section, exhibits an important breakup coupling effect on the elastic scattering, whereas $^8$B, with a large breakup cross section, shows a very modest coupling effect \cite{keeley1,a3}. Further investigation found that the coupling effect on the modulus of the S matrix, which is connected with changes in the imaginary part of the potential, is almost negligible for $^8$B and largest for $^6$Li. By contrast, for arg(S), which is connected with changes to the real part of the potential, the coupling effect is greatest for $^8$B, smallest for $^7$Be and intermediate for $^6$Li.  

In what follows,
 in Chapter II we include the experimental details, in chapter III the reduction of the data with subsections the elastic scattering channel, the breakup channel and other reaction channels, open at this energy. Finally in chapter IV, we give details of our theory and in chapter V we make a general discussion, closing up with a summary of our results.

\section{Experimental Details \label{sec:exp}}.

\begin{figure}
\begin{center}
\includegraphics[scale=0.45]{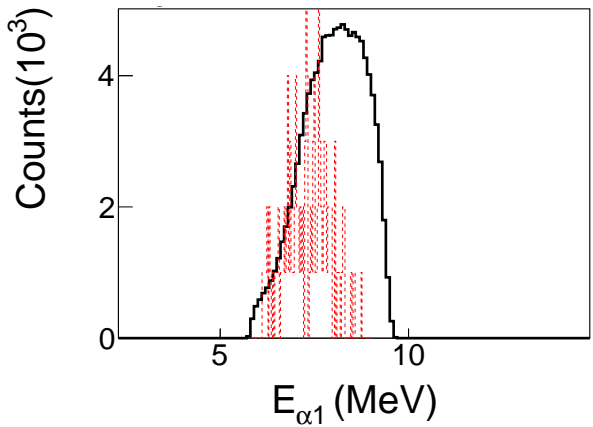}
\caption{Ungated alpha spectrum collected with MAGNEX magnetic spectrometer \cite{capp}. Alpha's were well discriminated from other particles via the  $\Delta$E - \ E technique, obtained via the focal plane gas and silicon detectors \cite{capp3}. With the dashed line (in red) is designated a simulated alpha spectrum \cite{sgouros,pakou12}, used to highlight the energy region where $\alpha$- particles are expected due to breakup. This is arbitrary normalized to the data.  Obvsiously other $\alpha$ - particles are also detected by MAGNEX, originating via other reaction processes  - see text. }\label{fig:alpha1}
\end{center}
\end{figure}
\begin{figure}
\begin{center}
\includegraphics[scale=0.45]{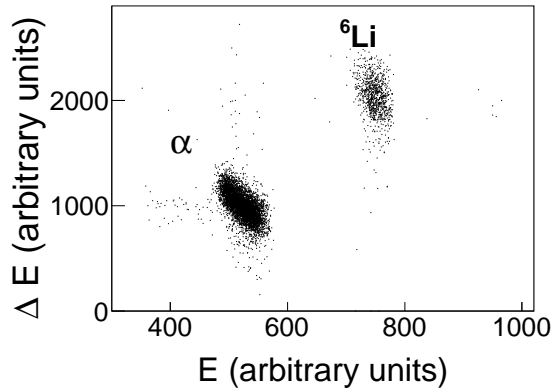}
\caption{A $\Delta$E - E spectrum collected by the focal plane detector of MAGNEX. }\label{fig:magnex}
\end{center}
\end{figure}

\begin{figure}
\begin{center}
\includegraphics[scale=0.45]{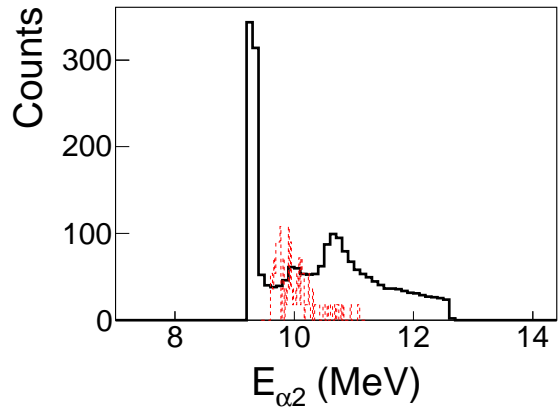}
\caption{Ungated particle spectrum obtained in $\Delta$E, the first DSSSD stage of the telescope module of  GLORIA array \cite{gloria}.  With the dashed line (in red) is designated a simulated  alpha spectrum \cite{sgouros,pakou12}, used to highlight the energy region where $\alpha$- particles are expected due to breakup. Obvsiously other $\alpha$ - particles are also detected by the telescope, originating via other reaction processes  - see text.The energy spectrum is transformed as being $\alpha$ - particles before the Ta mask. Also this is arbitrary normalized to the data.}\label{fig:alpha2}
\end{center}
\end{figure}

The experiment was performed at the MAGNEX facility of the Istituto Nazionale di Fisica Nucleare Laboratori Nazionali del Sud (INFN - LNS) 
in Catania, Italy, in inverse kinematics. For elastic scattering the MAGNEX magnetic spectrometer \cite{capp} was used, and the elastically scattered heavy ejectile, $^9$Be, was observed. MAGNEX was operated  at one position  with the optical axis at $\theta_\mathrm{opt}=6^\circ$, covering an angular range between 2.5$^\circ$ to 12$^\circ$. The combination of inverse kinematics and MAGNEX, which can operate at angles very close to zero, ensured an angular range for the angular distribution of elastic scattering, between $\theta$$_{c.m.}$ $\approx$ 25 $^\circ$ to 150$^\circ$ by changing only two magnetic fields. 

For the breakup measurement we followed the technique applied before in similar experiments \cite{pakou5,pakou6,pakou11,pakou12} with MAGNEX, used to detect one of the $\alpha$'s in an almost full angular range namely $\theta$$_{lab}$ = 2.5$^\circ$ to 12$^\circ$. The elastically scattered $^9$Be ions were most of them swept out by appropriate magnetic fields, allowing the detection of alphas in the energy range 6 to 9.5 MeV corresponding, according to our simulation \cite{sgouros,pakou12}, to the full energy phase space (see Fig. \ref{fig:alpha1}). The energy spectrum produced by the simulation is indicated with the dashed lines (in red). A comprehensive description of our simulation , as applied in this work, is given in Ref. \cite{pakou11}. It can be seen in Fig. \ref{fig:alpha1}, that $\alpha$ -particles of other origin are also present. These are due to a charge exchange reaction $^9$Be + p $\rightarrow$ $^9$B + n.  We will come back to this point also later on. A possible remainder, if any, of elastic scattering was rejected off line by the appropriate cuts in two dimensional $\Delta$E vs.\ E spectra, obtained from the focal plane gas and silicon detectors \cite{capp3} (Fig. \ref{fig:magnex}). A typical spectrum is shown in Fig. \ref{fig:magnex}, taken by the gas detector and one of the silicon detectors.
\begin{figure}
\begin{center}
\includegraphics[scale=0.45]{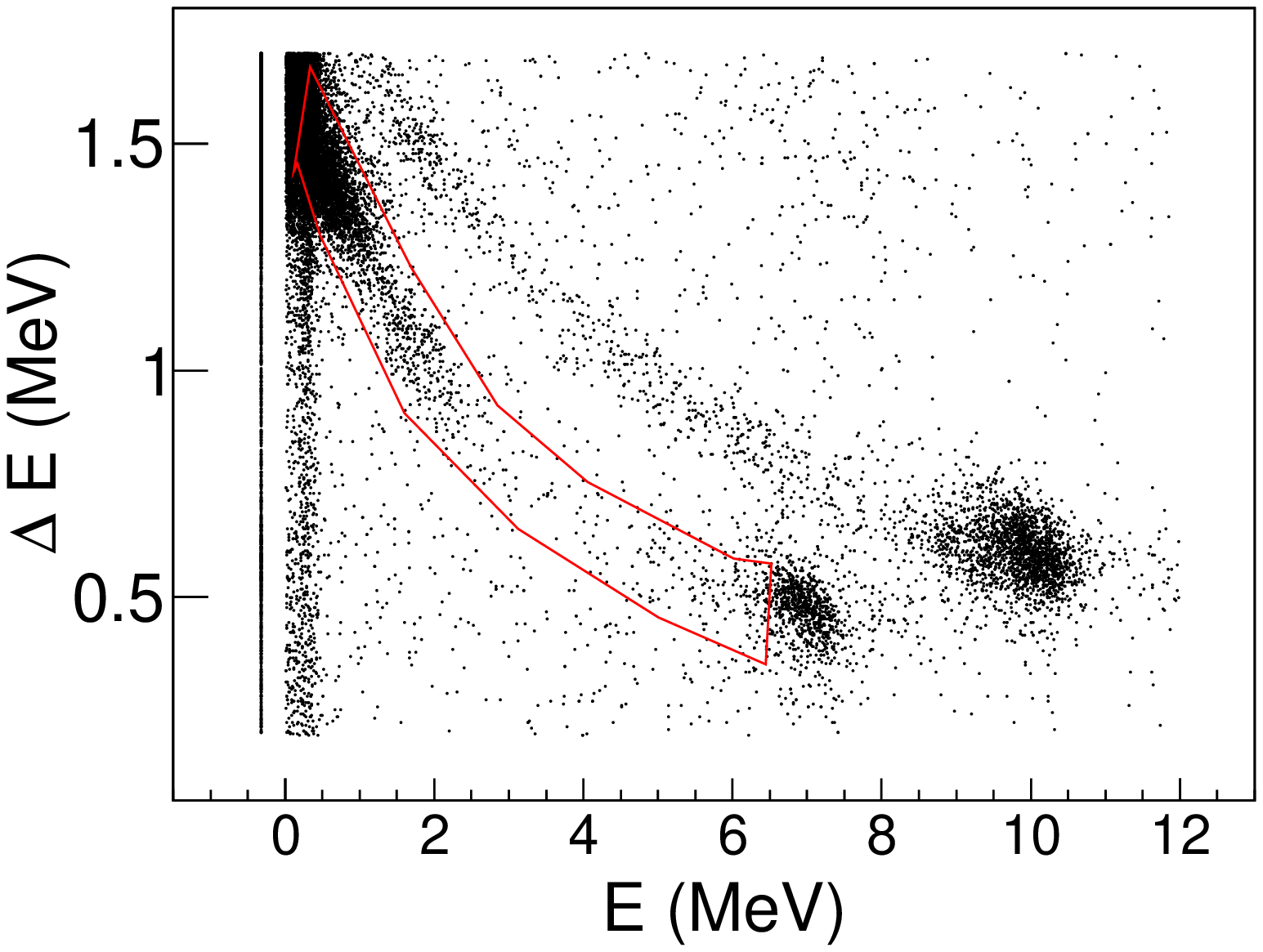}
\caption{A $\Delta$E - E spectrum collected by the module of the GLORIA array and one of the 16 strips at $\approx$ 23.9$^\circ$. The contour of recoiling protons taken in coincidence with $\alpha$ - particles in MAGNEX and $\Delta$E stage of GLORIA telescope, is designated with the solid line (in red).}\label{fig:bigloria}
\end{center}
\end{figure}

\begin{figure}
\begin{center}
\includegraphics[scale=0.35]{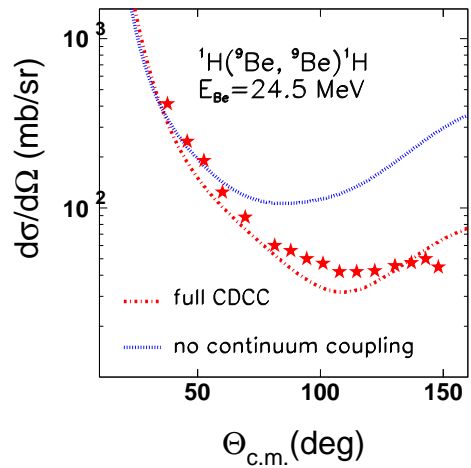}
\caption{Elastic scattering angular distribution for $^9$Be + p at 24.5 MeV (middle of the target).  Our data are compared with CDCC calculations. Data statistical errors are included in the size of the symbols (see text).}\label{fig:low}
\end{center}
\end{figure}
\begin{figure}
\begin{center}
\includegraphics[scale=0.35]{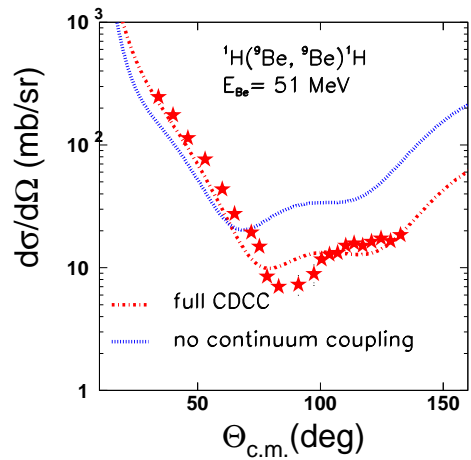}
\caption{Elastic scattering angular distribution for $^9$Be + p at 51 MeV. Previous data  \cite{pakou9,pakou10}, are compared with present CDCC calculations. Data statistical errors are included in the size of the symbols (see text).}\label{fig:high}
\end{center}
\end{figure}

The second $\alpha$ - particle was stopped in the first stage of a $\Delta$E - E silicon module of the telescope array GLORIA \cite{gloria}. The module comprised of  a $\Delta$E  Double Sided Silicon Strip detector (DSSSD) and an E pad  with thicknesses of 40 $\mu$m and 500 $\mu$m respectively. It was located, downstream of the beam, 125 mm far away from a CH$_{2}$ target, 492 $\mu$g/cm$^2$ thick, allocating an angular range of $\theta$$_{lab}$ = 5.8$^\circ$ to 28.5$^\circ$ . It was masked by a tantalum foil 18.2 $\mu$m thick, for preventing a deterioration of the detector material due to the strong Rutherford elastic scattering.  A particle spectrum collected with the DSSSD $\Delta$E module of the GLORIA telescope is presented in Fig. \ref{fig:alpha2}. The energy part, designated with the dashed line (in red) is the part of $\alpha$ - particles due to breakup in an $\alpha$ - $\alpha$ - p coincidence condition, as simulated by our code MULTIP \cite{sgouros}. The steep peak on the left is mainly due to protons and deuterons as well as alphas of various origins. In our event by event code all these particles were rejected with a restriction for particles above the energy of 9.1 MeV. The strong peak on the right, is identified through our simulation to originate via the stripping reaction $^9$Be + p $\rightarrow$ $^8$Be + d. This part is also avoided in our event by event analysis due to an $\alpha$ - $\alpha$ - p triple coincidence requirement. 

Finally the recoiling protons are observed and identified via the $\Delta$E - E technique by the telescope of the GLORIA array \cite{gloria}, used in this experiment. A bi - dimensional spectrum is shown in Fig. \ref{fig:bigloria}. Likewise deuterons, produced in the stripping reaction $^9$Be + p $\rightarrow$ $^8$Be + d, are resolved and identified in the same detector - see
Fig. \ref{fig:bigloria}.

\section{Data reduction\label{sec:reduction}}

\subsection{Elastic scattering}
The data reduction technique is similar to that adopted in a series of experiments performed at the MAGNEX facility with weakly bound light nuclei as $^{6,7}$Li and $^9$Be on a proton target \cite{pakou3,pakou4,pakou5,pakou6}, and it is mainly based on ray reconstruction for an accurate determination of the scattering angle and energy \cite{capp2,capp8,capp9,man}. For an angular step of $\sim$0.5$^\circ$, the counts were integrated and the solid angle, defined by 4 slits located at 250 mm from the target, was calculated taking into account the contour of the reconstructed ($\theta$$_i$, $\phi$$_i$) locus \cite{capp7}. The solid angle uncertainty is estimated to be $\sim$ 2\%. The beam charge was collected by a Faraday cup set at the entrance of MAGNEX. Taking into account the beam flux, the scattering centers of the CH$_2$ target with thickness 303$\mu$g/cm$^2$, the differential cross section angular distribution was determined and presented in Fig. \ref{fig:low}. Our previous data at 5.67 A MeV,  obtained under similar experimental conditions  \cite{pakou9,pakou10} will be considered in this work for comparison reasons and are presented in Fig. \ref{fig:high}. Uncertainties included in the data points of Figs. \ref{fig:low} and \ref{fig:high} are 1 to 3 \% and are due to statistical errors. This uncertainty will increase to $\approx$ 12 \% if we take into account systematic errors in flux, target and solid angle as 5, 10 and 2 \% respectively.

\subsection{The Breakup}
\begin{figure}
\begin{center}
\includegraphics[scale=0.45]{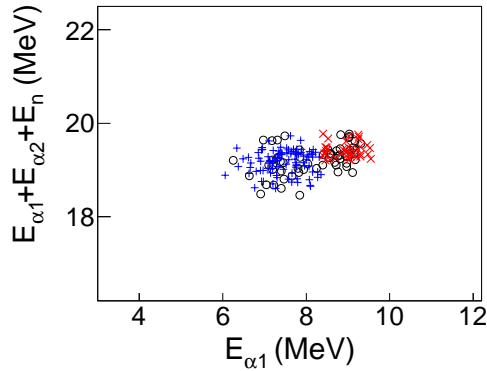}
\caption{Reconstructed sum  $\alpha$ + $\alpha$ + n energy versus the energy of an individual $\alpha$ - particle  detected in MAGNEX. Data are designated with the open circles, and simulation related with breakup and charge exchange - see text- with crosses (in blue) and x's (in red) respectively. It should be noted that our experimental setup was optimized on breakup, therefore the acceptance for the charge exchange reaction was small and only few events are observed for this reaction. }\label{fig:sum}
\end{center}
\end{figure}

\begin{figure}
\begin{center}
\includegraphics[scale=0.45]{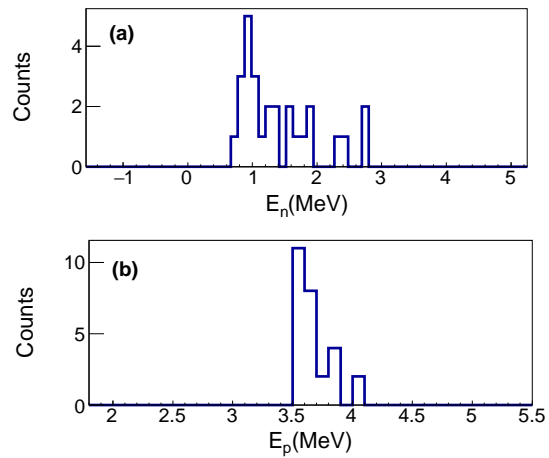}
\caption{Reconstructed energy spectra for (a)the unobserved neutron (b) the recoiling proton. }\label{fig:energy}
\end{center}
\end{figure}

The reduction of the data is based on a full kinematics approach. For that, an event by event code was developed with initial conditions on: (a) $\alpha$ - particle gates  and (b) proton recoil gates. Alpha's on MAGNEX are well discriminated from other particles, as it was already said, via $\Delta$E - E techniques (Fig. \ref{fig:magnex}). The gate on MAGNEX $\alpha$ - particles, related with the breakup events  is indicated  by our simulation, shown in Fig. \ref{fig:alpha1}. Later on, this was validated by the reconstructed sum plot shown in Fig. \ref{fig:sum}. This gate is then taken for $\alpha$ - particles with energies between 6 to 8.6 MeV. The other $\alpha$ - particle is stopped in the first stage of the $\Delta$E - E DSSSD telescope. A gate for particles with energy 9.1 MeV is required here (Fig. \ref{fig:alpha2}), to reject proton - deuterons and $\alpha$ - particles of other origin. Finally proton gates, defined by contours to recoiling protons through $\Delta$E, well resolved from deuterons, are additionally taken into account (see Fig. \ref{fig:bigloria}). A clear signature of a breakup event is then searched, obeying to a triple coincidence requirement and tagged by energy and angle. When such an event is found, the energy and angle spectrum of the unobserved event, here the neutron, can be reconstructed by applying the momentum
conservation law. In Fig. \ref{fig:energy}a, we present a neutron energy spectrum while in  Fig. \ref{fig:energy}b, the corresponding one for the recoiling proton.  Relative spectra as well as a Q - value spectrum can be also reconstructed. The last is given by the following equation

 \begin{equation}
   Q= E_{\alpha1} + E_{\alpha2} + E_n + E_p  - E_{beam} = E_{tot} - E_{beam}
\end{equation}

where, E$_{\alpha 1}$ is the kinetic energy of the alpha particle detected in MAGNEX, E$_{\alpha 2}$ is the kinetic energy of the alpha particle detected in the first stage of the EXPADES telescope where it stops, E$_n$ is the kinetic energy of the undetected neutron, E$_p$ is the kinetic energy of the recoiling proton identified in GLORIA module and E$_{beam}$ is the middle target beam energy. 

A Q - value spectrum is presented in Fig. \ref{fig:q}a. We can see that the peak concentrates at an energy of   Q $\approx$ -1.57 MeV, which is the separation energy of $^9$Be to $\alpha$ + $\alpha$ + n, therefore we consider that the events in our E$_p$ spectrum in Fig. \ref{fig:energy}b are the breakup events. Taking into account the yield of the E$_p$ spectrum, the beam flux and the efficiency of our system, the last determined via our simulation \cite{sgouros,pakou12}, a breakup cross section is determined as $\sigma$$_{break}$ = (2.5 $\pm$ 1) mb. The error includes the statistical error, 12 \%, and a 10 \% error in the efficiency determination. However the major part of the error is due to the energy loss and straggling of $\alpha$ - particles in the Ta foil \cite{stop}, which can change our $\alpha$$_2$ - particle gate restriction in $\Delta$E spectrum of the silicon telescope and further on an error in the definition of the $\alpha$$_1$ energy window defined for the MAGNEX $\alpha$ - spectrum.

\subsection{The p($^9$Be, n)$^9$B reaction}

\begin{figure}
\begin{center}
\includegraphics[scale=0.45]{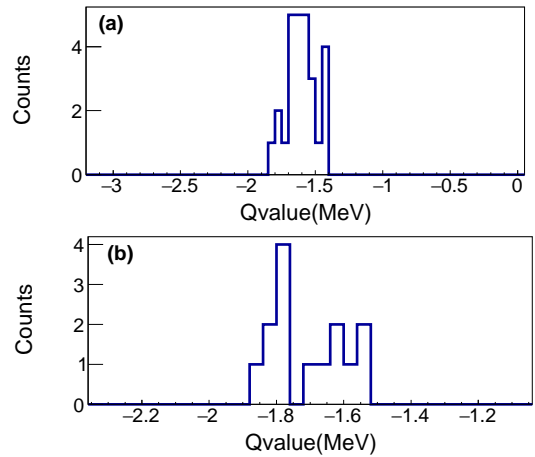}
\caption{Reconstructed Q - value spectra (a)With Ea2 gate $\geq$ 9.1 MeV and Ea1 $\leq$ 8.5   (b) With Ea2 gate $\leq$  9.1 MeV and Ea1 $\geq$ 8.5. See also Fig. \ref{fig:alpha1}, \ref{fig:alpha2} and \ref{fig:sum}. }\label{fig:q}
\end{center}
\end{figure}

An inspection of Fig. \ref{fig:q}a, discloses the presence of very few events with Q $\approx$ -1.85 MeV, the Q - value of a charge exchange reaction $^9$Be + p $\rightarrow$ $^9$B + n. This ignited an analysis with our event by event code but for particles in the $\Delta$E stage of the GLORIA module (Fig. \ref{fig:alpha2}) below 9.1 MeV. Defining a gate for $\alpha$$_2$ at lower and lower  energies in $\Delta$E, a well defined Q - value peak starts to develop around -1.85 MeV. This is demonstrated in Fig. \ref{fig:q}b, where the charge exchange Q - value peak is very well resolved from a Q -value peak associated with the breakup (Q = -1.57 MeV). Subsequently we have run our event by event code specifically for the charge exchange reaction but with a new gate for E$_{\alpha 2}$ and a gate in the Q - value. Taking into account the efficiency of our system for the charge exchange reaction ( the simulation was adjusted for the charge exchange reaction) the beam flux and the scattering centers we have determined a cross section of $\sigma$$_{CE}$ = 110$\pm$ 40mb. Previous data with full angular distributions exist in the literature at similar energies, literary speaking at 2.56 A MeV, 2.6 A MeV  and 2.9 A MeV \cite{rea2}. These data were integrated and interpolated to our energy at 2.72 A MeV with a reaction cross section equal to $\sigma$$_{CE}$ = 98 mb. This value is in very good agreement with the present result.

\subsection{The p($^9$Be, d) reaction }

\begin{figure}
\begin{center}
\includegraphics[scale=0.4]{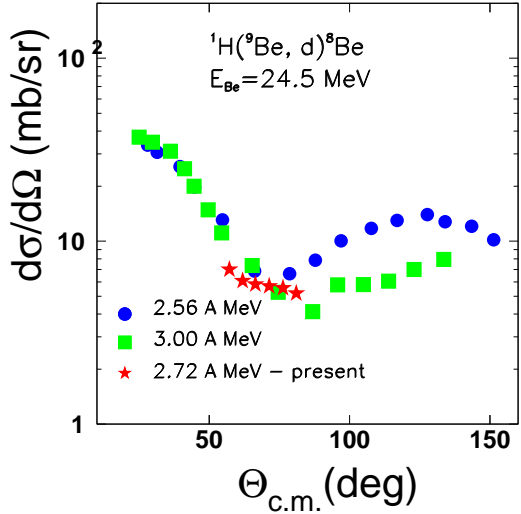}
\caption{Angular distributions of the reaction p + $^9$Be  $\rightarrow$ d + $^8$Be. Previous data \cite{rea1} at 2.56 and 3 A MeV  are designated with the filled circles and boxes respectively (in blue and green) while the present with the stars ( in red) at 2.72 A MeV. Data errors are included in the size of the symbols.}\label{fig:repd}
\end{center}
\end{figure}

Previous experimental data exist in the literature for this (p,d) reaction at 2.56 and 3 A MeV \cite{rea1}. Full angular distributions reported in \cite{rea1} have been included in Fig. \ref{fig:repd}. Our present analysis was performed through the deuteron products taking into account single spectra from the silicon telescope. Due to a high counting rate at the most forward angles the resolution between the deuterons two body spot in the $\Delta$E - E plots and elastic proton spot was not good, therefore the analysis was limited in a narrow angular range between 23 to 33 degrees, that is the most backward strips of our silicon detector. Our results are compared very well with the previous data, as it is shown in Fig. \ref{fig:repd}. Therefore the previous angular distributions were integrated and interpolated to our energy with a reaction cross section equal to $\sigma$$_{p,d}$ = 161 $\pm$ 60 mb.
\begin{figure}
\begin{center}
\includegraphics[scale=0.4]{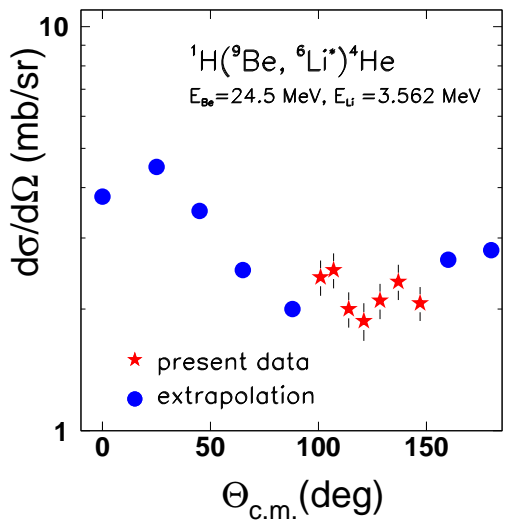}
\caption{Angular distributions of the reaction p + $^9$Be  $\rightarrow$ $\alpha$ + $^6$Li$^*$ feeding the 3.562 MeV excited state in $^6$Li. With stars are designated the present results. Due to the limited angular range, an extrapolation was attempted according to the angular distribution shape of ground state transition (see text). }\label{fig:reli}
\end{center}
\end{figure}
\subsection{The p($^9$Be, $^6$Li) reaction}

Previous experimental data exist also in the literature for this  reaction at 2.56 and 3 A MeV \cite{rea1,rea3,rea4}. It seems that the reaction goes through an intermediate nucleus, $^{10}$B, which decays to $^6$Li and an $\alpha$ - particle,  populating besides the  ground state of $^6$Li, the first two resonances at 2.186 and 3.562 MeV. Full angular distributions exist only for the ground state transition. These results were integrated and interpolated for our energy to a value of $\sigma$$_{6Li}$ = 167 $\pm$50 mb.
Cross section data for the first resonance at 2.186 MeV do not exist in the literature for our energy. An indication, is given in Ref. \cite{rea1} with ratio  of the cross sections at $\theta$$_{lab}$ = 35 $^\circ$ between ground state of the (p, d) reaction and first excited state of the (p, $^6$Li) reaction to be almost four. In this respect, a very approximate determination of the cross section will be $\sigma$$_1$ $\approx$ 40 $\pm$ 20 mb. The feeding of $^6$Li via the  metastable state at  3.562 MeV was studied by the authors of Ref. \cite{rea3,rea4} but only for proton energy at 2.56 MeV, where a steep resonance exist, and these data can not be considered here.  In the present experiment due to the particular setup acceptance only the  feeding via this case of 3.562 MeV state has been seen and only in a limited angular range. The measurement was performed by observing the recoiling $^6$Li, detected in MAGNEX (see Fig. \ref{fig:magnex}). A new ray reconstruction of the data with gate to lithium reaction products  was performed and an angular distribution but only with backward data is obtained and shown in Fig. \ref{fig:reli}). These data  had to be extrapolated to forward and backward angles. This was done assuming the shape of the ground state angular distribution. Therefore this reaction cross section, obtained by integration of this angular distribution is given with caution as $\sigma$$_2$ = 33 $\pm$15 mb, assuming that this is an upper limit. Also with caution we give  a total reaction cross section for this reaction channel as $\sigma$ $\approx$ 240 $\pm$ 55 mb.

\section{ Four - body  CDCC calculation}

We studied the $^9\text{Be}+p$ reaction at two energies, namely 2.72 A MeV and 5.67 A MeV, using four-body Continuum-Discretized Coupled-Channel (CDCC) calculations, which include the coupling to breakup channels explicitly. The $^{9}$Be projectile states were generated using a pseudostate approach, the analytical Transformed Harmonic Oscillator (THO) method presented in Ref.~\cite{ro1}. Within this framework, continuum states are obtained by diagonalizing the three-body Hamiltonian in a THO basis, and may contain resonant and non-resonant components. The method has been previously applied to $^9$Be-induced reactions on heavier targets~\cite{ro2,ro3}. In the present case, we need $p$-$^4$He and $p$-$n$ potentials to generate the corresponding form factors. For the $p$-$n$ system, we employed the Gaussian potential of Ref.~\cite{Austern}. For $p$-$^4$He, we fitted optical potentials to describe the elastic scattering data reported in Refs.~\cite{Kreger,Miller}. We have included real and imaginary Woods-Saxon volume  terms with $r_0=1.1$ fm and $a=0.477$ fm. At the lowest energy, an additional spin-orbit term was needed to achieve the best fit of the $p$-$^4$He data. However, the multipole expansion used to generate the CDCC form factors~\cite{MRG08} does not allow spin-orbit couplings. Therefore, calculations were complemented with a global $^{9}\text{Be}+p$ spin-orbit term following the Watson prescription~\cite{Watson}. With these ingredients, we solved the coupled-channels problem including projectile states up to $j^\pi=5/2^\pm$, which were coupled to all multipole orders. Convergence of the elastic and breakup cross sections was obtained by fixing a maximum continuum energy of $E_{max}=8$ MeV and a total angular momentum (including the relative motion and the spin of the proton) of $J_{max}=20$. It is worth noting that, once the structure model for $^9$Be and the optical potential are fixed, the CDCC calculations do not involve any parameter fitting and provide absolute cross sections to compare with the experimental data. Our elastic scattering results  for the energies of 2.72 and 5.67 A MeV, are compared in an excellent way with the data in Figs.~\ref{fig:low} and~\ref{fig:high}, respectively. Computed breakup and total reaction cross sections are compared with experimental values in Table I and found in good agreement. In this respect, our three-body model for $^9$Be is well established and able to describe the scattering on heavy and light targets at different near-barrier energies. Such a model is relevant for the calculation of the corresponding radiative capture reaction rate, reported in Ref. \cite{ro1}, so the present results also support this estimation.

\begin{table}

\caption{Comparisons of experimental (e) and four \ - body CDCC calculations (t) of breakup cross sections (BU),  and of total reaction cross sections (tot) at 2.72 A MeV and 5.67 A MeV}

\begin{ruledtabular}
\begin{tabular}{ccccc}

E(A MeV)&$\sigma$$^{e}$$_{BU}$ (mb)&$\sigma$$^{t}$$_{BU}$ (mb)&$\sigma$$^{e}$$_{tot}$ (mb)&$\sigma$$^{t}$$_{tot}$ (mb)\\

2.72&2.5$\pm$1&3.67&510$\pm$90&433\\
5.67&142$\pm$20&163&&820

\end{tabular} 

\end{ruledtabular}
\end{table}

\section{Discussion - Summary}

We have obtained new results for $^9$Be + p elastic scattering and breakup at 2.72 A MeV in inverse kinematics. The elastic scattering was performed detecting the heavy ejectile, $^9$Be, at the magnetic spectrometer MAGNEX. For breakup, the measurement was performed in a full kinematics approach within a triple $\alpha$ - $\alpha$ - p coincidence approach, by detecting one of the $\alpha$ - particles in MAGNEX and the second $\alpha$ - particle as well as the proton recoil in a silicon telescope of the GLORIA detection array. Additionally, some results were obtained  for the reactions $^9$Be + p$\rightarrow$ $^8$Be + d, $^9$Be + p$\rightarrow$ $^9$B + n and  $^9$Be + p$\rightarrow$ $^6$Li$^*$ + $\alpha$, adequate to validate  the first two reactions, but not the third one.  In all three reactions,  previous comprehensive angular distribution data exist and were integrated giving us the possibility to estimate a total reaction cross section, $\sigma$$_{tot}$= 510 $\pm$ 90 mb. We should note that this cross section is given with caution, since for one of the reactions we were not be able to validate the previous results. Our new $^9$Be + p results combined with the previous at 5.67 A MeV, enables a global description of the $^9$Be + p  continuum. Our results for elastic scattering and breakup are included in Figs. \ref{fig:low}-\ref{fig:high} and Table I, respectively.

As it is shown in  Figs. \ref{fig:low}-\ref{fig:high}, where calculations with no coupling and coupling to continuum is presented in a four- body CDCC approach,  the coupling effect to continuum is strong. In particular the effect for the lowest energy constitutes a spectacular result, indicating that in certain nuclei depending on the breakup mechanism, even if the breakup cross section is small (see Table I) the coupling can be very strong. Same conclusions were drawn before for both $^6$Li and $^7$Li projectiles \cite{pakou5,pakou6} in accordance with other findings \cite{keeley1}.

On  the other hand, the primary goal of this research was to unfold step by step the various mechanisms of the breakup process, studying the $^9$Be + p system at various energies.  Our calculations at the two energies, namely 2.72 and 5.67 A MeV, indicate that the reaction mechanism differs between the two energies. At the lower energy, we obtain a small breakup cross section of $\sigma_{\rm BU}=3.67$ mb, in very good agreement with the experimental value of (2.5 $\pm$ 1) mb, which  proceeds through near-threshold $1/2^+$ state in $^9$Be. The available energy here is  just 0.88 MeV above threshold which means that mainly the resonance at 1.684 (1/2$^+$) contribute to this coupling. At the higher energy, in contrast, at least $\sim 50$\% of the total breakup cross section of $\sigma_{\rm BU}=163$ mb, is carried by states around the narrow 5/2$^-$ resonance in $^{9}$Be, in accordance with the experimental value of (142 $\pm$ 20) mb. Our three-body structure calculations for this resonance yield a wave function dominated by $L=2$ components in the relative $\alpha$-$\alpha$ motion, and $p_{3/2}$ components in the $\alpha$-$n$ subsystem. This might be consistent with an $\alpha + {^5\text{He}}$ breakup mode, since our resonance calculations favor configurations with $^{5}$He in its ground state. We should underline here that the experimental breakup results indicate a major contribution via the $^5$He + $^4$He configuration, but we should also note that our three-body calculations for $^{9}$Be contain, in principle, all these configurations on an equal footing, thus a quantitative separation of the different breakup modes is not trivial for a direct comparison with the experiment. The CDCC calculations provide also total reaction cross sections as it can be seen from Table I. The present measurement at low energy provide some results capable for validating more comprehensive previous results existing in the literature. The last include angular distributions, which were integrated summed up and shown in Table I, in very good consistency with the calculation. For the higher energy our setup did not allow the extraction of data for other reactions. Therefore no further evaluation of previous data was considered.
 
 In summary
 
 The elastic scattering and breakup as well as other involved reaction channels were measured for $^9$Be + p in inverse kinematics at 2.72 A MeV and were considered in a four - body CDCC theoretical framework. Likewise data at 5.67 A MeV were also considered under the same footing. An excellent consistency was observed between experiment and theory concluding the following
 \begin{enumerate}
 \item 
 The system $^9$Be + p can be described in an excellent way in a four -  body CDCC framework at low energies as   2.73 A MeV and 5.67 A MeV. As a consequence, our three-body model for $^9$Be is further validated and therefore well established to be used for relevant radiative capture reaction rates.
 \item
 
 The non reciprocity between the strengths of breakup and the couplimg to continuum is apparent, within this system and the low energy of 2.73 A MeV. While the breakup cross section is small and negligible  in comparison with other more favored reaction channels, coupling to continuum is strong and essential in reproducing elastic scattering data. 
 
 \item
 The breakup probability drops sharply from the higher to lower energy by a factor of $\approx$ 50, while other reaction channels only by a factor $\approx$ 2.
 
 \item
 According to our theory, at the low energy, breakup can be interpreted to be produced by direct or/and sequential processes through the 1.684 MeV (1/2$^+$) resonance through the $\alpha$ + $\alpha$ + n configuration, while at the higher energy by far the contribution is due to the 2.43 MeV (5/2$^-$) resonance and the $^5$He + $^4$He configuration. These findings are in good compatibility with the data.
 
\end{enumerate}
\section*{Acknowledgments}

The research leading to these results has received funding from the European Union HORIZON2020 research and innovation programme under Grant Agreement n 654002 - ENSAR2. Also was partially supported by the European Research Council (ERC) under Grant Agreement No. 714625, and by the Ministry of Science, Innovation and Universities of Spain, Grant No.\ PGC2018-095640-B-I00, and by the Spanish Ministry of Economy and Competitiveness, the European
Regional Development Fund (FEDER), under Project No.
FIS2017-88410-P, and by SID funds 2019 (Universit\`{a} degli Studi di Padova, Italy) under project No.~CASA\_SID19\_01.

\end{document}